# A Comprehensive Assessment Strategy for Physics Laboratory Courses

Rajesh B. Khaparde
Homi Bhabha Centre for Science Education
Tata Institute of Fundamental Research
V. N. Purav Marg, Mankhurd, Mumbai 400088, INDIA

Email: rajeshkhaparde@gmail.com

**Abstract**
The objective of physics laboratory training is to develop, in students, a variety of important cognitive and psycho-motor abilities related to experimental physics. These include conceptual understanding, procedural understanding, experimental skills and the experimental problem solving ability. It has been noted that strategies adopted for the assessment of what students learn and develop through a laboratory course are often inconsistent with the objectives of the laboratory courses. The author has developed a comprehensive assessment strategy which can be used at the school, college and university level. The strategy is based on four tools of assessment, namely, test on conceptual understanding, test on procedural understanding, an experimental test, and the continuous assessment. The relative weightage for each of the four tools depends on the level and emphasis of the laboratory course. The four tools of assessment, with respect to the type of questions, design, grading schemes, administration of each tool have been described with a few sample questions for each tool of assessment. This assessment strategy is being practiced and its effectiveness studied during a series of special courses in experimental physics in India. Furthermore, a survey was carried out with university teachers from across India to check the acceptability and feasibility of using this strategy for larger number of students in universities and most of the participating teachers supported the use of this strategy.

**Introduction**
Assessment of laboratory performance has always been a matter of concern and discussions during faculty meetings in colleges and universities. However, it has not been given its due importance by the Physics Education Researchers and university administrators as far as developing and implementing novel strategies of assessment are concerned. A few researchers have reported noteworthy developmental work on laboratory performance test and strategies for physics laboratory courses. (Wall 1951; Kruglak 1954; Kruglak 1958; Moreira 1980; Theysohn 1983) In this paper, a comprehensive assessment strategy for physics laboratory courses is described, which the author had developed as part of a major project on development of 'contents' and 'strategies' for training in experimental physics. (Khaparde, Pradhan, 2009).

It is well accepted that physics laboratory courses are supposed to develop in students important cognitive, psycho-motor, attitudinal and affective abilities related to experimental physics, which essentially include, conceptual understanding, procedural understanding, experimental skills and experimental problem solving ability. In accordance with these major objectives, the assessment strategy should have suitable assessment of all four aspects with appropriate weightage for each and the final grading should be only on the basis of the convergence of the results obtained through such multiple tools of assessment.

The assessment strategy presented below is based on four tools of assessment namely i) a test on conceptual understanding, ii) a test on procedural understanding, iii) an experimental test and iv) continuous assessment which is based on students regular laboratory work throughout the course. The first two are paper and pencil type tests, which correspond to a single variable each, namely, the level (or score) of a student's conceptual understanding and the procedural understanding, respectively. The third focuses around the variable 'experimental skill', but is not a single variable measure; rather it is a composite measure of the experimental skills, problem solving ability, conceptual understanding, and procedural understanding. It is difficult to isolate the contributions of each of the four variables to this composite measure. In the following subsections various tools of assessment referred to above are described in details.

**Test on Conceptual Understanding**
Conceptual understanding is the understanding of concepts, their interdependence and ideas in science, which are based on facts, laws, and principles. According to this strategy, the first tool of assessment is a 'common for all' pen and paper type written test to quantitatively measure the students' conceptual understanding developed in students through a laboratory course. This test is essentially 'context linked', i.e., the questions had a direct link to or an application of the conceptual understanding involved in the experiments and demonstrations. The context or the situations around which the questions were framed were new and novel. This was essential to evaluate the extent to which the students had developed conceptual understanding through the given laboratory course.

This test had mainly 'multiple choice' questions. It also had 'match the pairs' and 'fill in the blank' type of questions. Some questions involved drawing or completing a figure or a schematic/ray diagram. The text in the questions was supported by schematics or figures to clearly explain the 'question' to the students. This test was validated and thoroughly discussed with the teaching faculty members. Each student was given a question paper. Students were asked to mark their answers on the question paper itself, which were collected after the test and hence no separate answer paper was necessary. Blank papers were provided to the students for carrying out rough work. Students were given ample time to answer the given set of questions. In the grading scheme, marks were assigned for the correct answer as per the difficulty level of the question. The marks allotted to each question and to the whole test itself were decided by the number of questions and the level of difficulty. The students were informed about the grading scheme during the tests. To illustrate the 'contents' of a test on conceptual understanding, a few sample questions on conceptual understanding are given below.

1) Water drops are freely falling vertically down from the nozzle of a tap at regular intervals. Which of the following picture is most appropriate to describe the line of falling drops?

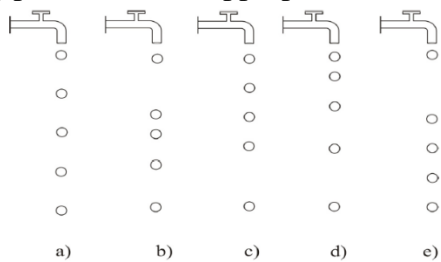

2) A student is given three laser sources which emit light of wavelengths $\lambda_1$, $\lambda_2$ and $\lambda_3$. When he illuminates a diffraction grating by these laser sources keeping the distance between the grating and the screen the same, he observes the given pattern of bright spots on the screen. What can you infer about the wavelengths $\lambda_1$, $\lambda_2$ and $\lambda_3$?

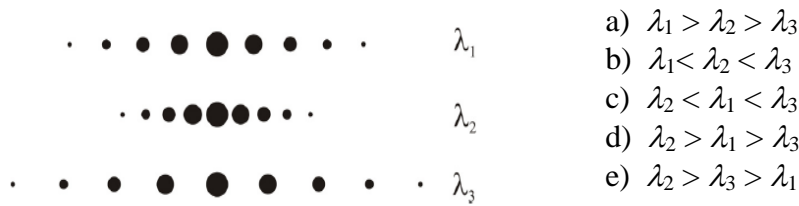

a) $\lambda_1 > \lambda_2 > \lambda_3$
b) $\lambda_1 < \lambda_2 < \lambda_3$
c) $\lambda_2 < \lambda_1 < \lambda_3$
d) $\lambda_2 > \lambda_1 > \lambda_3$
e) $\lambda_2 > \lambda_3 > \lambda_1$

3) Which of the following statement(s) you think is/are not correct ?

   a) An equi-potential surface is a surface composed of all those points having the same value of the potential.
   b) No work is involved in moving a unit charge around on an equi-potential surface.
   c) There is no potential difference between any two points on this surface.
   d) The electric field is normal to equi-potential surfaces.
   e) The electric field at any point on the equi-potential surface is the same in magnitude.

4) A transformer is designed with two coils primary and secondary, each of 500 turns of copper wire of the same cross section, wound on the central leg of an ordinary laminated iron core. The resistance of the wire used for primary coil is 2.9 Ω and the resistance of the wire used for secondary coil is 3.8 Ω. (The secondary coil is wound over the primary coil).

The secondary coil is left open and the primary coil is connected to a 230 V, 50 Hz power supply. The peak current passing through the primary coil is found out to be 200 mA. If the secondary coil is shorted, the current in the primary coil will

   a) increase
   b) decrease
   c) remain the same
   d) will be near to zero

5) The emf induced in a wire by its motion across a magnetic field does not depend upon.

   a) The length of the wire
   b) The diameter of the wire
   c) The material of the wire
   d) The orientation of the wire
   e) The magnetic field strength

**Test on Procedural Understanding**
According to Gott and Duggan (1995), procedural understanding is the 'thinking behind the doing' or the decision-making in designing and performing experimental activities. It is the understanding of a set of ideas or concepts of evidence related to the 'knowing how' of science and related to designing experiments, planning measurements, observations, analyzing, and interpretation of data. It was felt that the significance of procedural understanding is subsumed and therefore lost under the rubric of 'experimental skills' (Khaparde, Pradhan, 2002). Thus according to the strategy, the level of procedural understanding, which is a cognitive understanding in its own right, should be measured quantitatively by the students' performance in a separately designed 'common for all' test on procedural understanding. This being 'conceptual in nature' it was felt that a written test with appropriately framed questions based on experimental situations should be designed. It was noted that this type of test is not a replacement of an experimental test in which procedural understanding and various other abilities are synthesized to solve the given experimental problem.

In this test, there were 'descriptive', 'essay' type, 'multiple choice', 'fill in the blank' and 'match the pairs' type of questions. The questions were 'context free', i.e., these questions had no direct bearing on a set of experiments and demonstrations included in the laboratory course. This test was very carefully designed and validated. The test had questions on various aspects related to the how, why and what of the design, measurement and data handling, e.g., devising a procedure for measuring a particular parameter, choosing an appropriate instrument, understanding relationships in instruments, determining the range and accuracy required, exercising warnings, controlling parameters, changing parameters, measures to reduce errors, variable structure, choosing values, sampling the data, the intervals between the readings, reliability, and validity of data, data representation, etc.

Each student was given a question paper along with an answer paper. Students were provided blank papers for necessary rough work. The question paper, the answer paper, and the blank papers were collected at the end of the allotted time for the test. Students were given ample time to answer the given set of questions and were informed about the grading scheme during the test. On account of variation of content, length of expected answers and difficulty levels, it was necessary to give different weightage to different questions in the grading scheme. To work out the grading scheme, a detailed model answer for each question was prepared, analyzed, and accordingly the marks were allotted to each question.

To illustrate the 'contents' of a test on procedural understanding, a few sample questions on procedural understanding are given below.

1) A rectangular body of material with known density has a cubic cavity inside it. Design and explain an experimental method to determine the size of this cavity and locate its position.

2) You are given a spring, known masses and a meter scale. Suggest an experimental method to determine the mass of the spring.

3) In an experiment, you are supposed to study the variation of resistance of a thermistor with temperature. In this experiment, which instruments will you need? Which parameters will you keep the same? Which parameters will you change and the change in which parameters will you record?

4) In an experiment to determine the coefficient of viscosity of a liquid, the liquid is transferred through a siphon made of a plastic tube of uniform cross section from one cylindrical container to the other of identical cross section and height. The expression, which is to be used for the determination of viscosity is,

$$\eta = \frac{a^4 \rho g t}{(4R^2 L) \ln(H_0 / H)}$$

where, $\eta$ is the coefficient of viscosity of the liquid, $a$ is the radius of the inner cross section of the tube, $\rho$ is the density of the liquid, $g$ is the acceleration due to gravity, $t$ is the time, the liquid takes to flow through the siphon, corresponding to the difference of levels of the liquid $H$ in the cylinders, $L$ is the length of the tube ($\approx 2$m), $H$ is the difference in the liquid levels in the two containers, initial value of which is $H_0$ and $R$ is the inner radius of the cylindrical container.

In this experiment which quantities will you measure ? Explain your choice of instrument for the measurements involved in the above experiment. Which quantities will you measure more accurately and why ?

**Experimental Test**
According to the strategy, experimental skills and problem solving ability should be quantitatively measured through a separate experimental test. Hence, the third tool of assessment was an experimental test. It is noted that the experimental test gives a composite measure of students' conceptual and procedural understanding, experimental skills and problem solving abilities required to effectively solve the given experimental problem. However since all these components of the composite measure are integrated with each other, separating their individual effect is hardly possible in such a test.

An identical experimental test was given to all the students. This ensured that all the students were tested on the 'same ground'. The experimental test was carefully designed to involve experimental skills and experimental problem solving abilities. Every student was individually given the set of necessary apparatus, the question paper, the answer paper, blank papers, and graph papers. The question paper had the necessary details of the experimental test such as the objectives, apparatus, description of the apparatus, useful data and the statement of the experimental problem. The students were asked to report, all the steps they have devised and followed, measurements, data, data analysis, and interpretation in a systematic manner. They were asked to report practically every observation taken and every procedural step followed. This comprehensive reporting allowed the graders to grade the students' performance after they complete the test based on only the report in their answer papers.

In accordance with the strategy, the students' performance was graded entirely through their comprehensive reports, totally avoiding subjective judgments based on the 'observations' and 'interrogations' by the grader/examiner. This was an important aspect of the strategy of assessment of students' performance in an experimental test (this strategy is regularly being used in the International Physics Olympiads). This considerably reduces the subjectivity of assessment present in the traditional strategy. Thus, teachers or examiners present in the laboratory during the test are not supposed to interact with the students, unless a student needed to consult them for technical reasons.

The grading scheme was carefully designed to evaluate the students' performance with respect to various aspects and abilities, which the experimental test is supposed to evaluate. First the required data was collected, a model answer was prepared, and various stages of the solution of the experimental test were identified. Only then the relative weightage for different stages was decided. It was felt that students' reporting is also an important aspect and hence some marks were reserved for reporting and presentation of the laboratory work. Marks were allotted to the experimental skills (reflected from the discrepancies between the expected and the reported data), understanding and application of the necessary theory and concepts, collection, organization and analysis of data, and overall approach towards handling the given experimental problem. For example, the statement of the problem for the experimental test on 'efficiency of a light emitting diode' was "Design and perform the necessary experiment to a) show that the current generated in the photodiode is linearly proportional to the intensity of the light falling on its sensitive area, b) study the variation of the efficiency of an LED with the current passing through it, and c) determine the total radiant power emitted by the LED and calculate its maximum efficiency."

**Continuous Assessment**
It was felt that it is important to monitor and evaluate students work during the regular laboratory course. Student's report of each experiment should form an important component of the assessment. Thus the fourth tool of assessment was continuous assessment based on students work in the laboratory and their reports. Students were asked to record every procedural step they adopted during the experimental work, observations, method, detailed data, data analysis, final results, and inferences in their reports . It was necessary that students complete the report of a given experiment before proceeding to the next one. The teachers were asked to observe the students during the laboratory course. The teacher were expected to continuously grade and correct the report of each experiment and give detailed comments, feedback, and the marks. The final marks had a combination of all the individual scores. A substantial portion of the total marks were reserved for this 'continuous' assessment.

**Evaluation of the effectiveness of the strategy**
This assessment strategy is being practiced during training of students and teachers and its effectiveness was studied at HBCSE-TIFR Mumbai, India (http://www.hbcse.tifr.res.in) through a series of special courses in experimental physics. From the students grades and feedback, it was noted that this assessment strategy is comprehensive, objective, valid, and reliable as an achievement test, as compared to the traditional assessment strategy being used for laboratory courses in India. This strategy is being employed as a regular assessment strategy at University of Mumbai-Department of Atomic Energy-Centre for Excellence in Basic Sciences, Mumbai, India (http://www.cbs.ac.in) and a few other undergraduate institutions in India.

Furthermore, a survey was carried out with 45 university faculty members, to study the acceptability and feasibility of using this strategy for larger number of students in universities. Detailed opinions and suggestions were collected through a "Questionnaire on Assessment Strategy". Majority of the participating teachers reported that this was a much better strategy as compared to the strategy being used in Indian universities. The teachers supported the use of all the four tools of assessment presented in this strategy, except on the use of identical experimental test for all the students, as this would require a large number of identical experimental setups.

**Conclusions**

This comprehensive assessment strategy involving all the four tools of assessment with an appropriate weightage for each can be used in physics laboratory courses. The first two tools could be administered and evaluated centrally for all the students enrolled for the laboratory course. The relative weightage should depend on the level and objectives of the laboratory course. As an example, for an introductory laboratory course, one may have a) 25% of the total marks for test on conceptual understanding b) 20% of the total marks for test on procedural understanding c) 25% of the total marks for the experimental test, and d) 30% of the total marks on the continuous assessment.

It is essential to realize and appreciate the fact that the assessment strategy, based on which each student is assessed and given grades that appear in the final score, is a factor which directly or indirectly affects the importance and the effectiveness of the laboratory course.